\documentclass[preprint,samsmath,amssymb,floatfix,nofootinbib,aps,showpacs]{revtex4}
\usepackage{amssymb}
\usepackage{graphicx}

\begin{document}

\title{More on phase diagram of Laponite}
\author{B.~Ruzicka$^{1,2}$, L.~Zulian$^{2,3}$, G.~Ruocco$^{1,2}$}
\affiliation{$^1$ Dipartimento di Fisica, Universit\'a di Roma
"La Sapienza", P.zle A. Moro 2, I-00185 Roma, Italy\\
$^2$ INFM-CRS SOFT, Universit\'a di Roma "La Sapienza", P.zle A.
Moro 2, I-00185, Roma, Italy.\\
$^3$ Dipartimento di Fisica, Universit\'a di Perugia, v. A.
Pascoli, I-06123 Perugia, Italy}
\date{\today}
\begin{abstract}
The phase diagram of a charged colloidal system (Laponite) has
been investigated by dynamic light scattering in a previously
unexplored range of salt and clay concentrations. Specifically the
clay weight and salt molar concentrations have been varied in the
ranges $C_w=0.004 \div 0.025$, $C_s=(1\times 10^{-3} \div 5\times
10^{-3}$) M respectively. As in the case of free salt water
samples ($C_s \simeq 1\times 10^{-4}$ M) an aging dynamics towards
two different arrested phases is found
 in the whole examined $C_w$ and $C_s$ range. Moreover a transition
 between these two different regimes
  is found for each investigated salt concentration. It is clear from these measurements
  that a revision of the
 phase diagram is necessary and a new "transition" line between two different
 arrested states is drawn.

\end{abstract}
 \maketitle

\section{Introduction}
In spite of intensive research on Laponite clay suspensions in
water, motivated by its intriguing properties and important
industrial applications, a general agreement about its phase
diagram is still lacking. It is however clear that a very rich
phenomenology appears varying clay concentration ($C_w$) and/or
salt concentration ($C_s$) in the system. Broadly speaking the
phase diagram \cite{Mourchid} shows that at relatively low salt
concentration, increasing clay weight one can obtain a liquid
phase, a gel (or glass) phase and a nematic gel, at high salt
concentrations one instead observes flocculation.
 The most
studied part of the phase diagram is the one at low salt
concentration (usually samples are prepared in pure deionized
water, $C_s \simeq 1 \times 10^{-4} M$) and intermediate clay
concentration ($C_w\simeq 3 \%$). In this region several dynamical
light scattering measurements have shown that the samples
initially liquids age towards an arrested phase \cite{Kroon1,
Bonn, Knaebel, Abou, Bellour} but the mechanism of aggregation is
not clear. The gel/glass state has been in fact attributed by
different authors to Wigner glass transition \cite{Bonn},
frustrated nematic transition \cite{Mourchid, Kroon, Gabriel,
Mourchid1}, micro-segregation \cite{Mourchid, Pignon2, Martin},
gelation \cite{Nicolai1, Nicolai2,Mongondry1}, etc.. Moreover
recent investigations have shown that also the low salt/low clay
concentration region of the phase diagram is very interesting and
that this is not a region of stable liquid phase as previously
supposed. In particular, by dynamic and static light scattering
measurements, an arrested phase has been surprisingly found also
at low clay concentrations for very long waiting times
\cite{Nicolai1,Nicolai2,Ruzicka1,Ruzicka2,Mongondry1}. Therefore
it is clear that the previously proposed phase diagrams
\cite{Mourchid, Tanaka1, Tanaka2} need to be revised also in the
low clay/low salt concentration region \cite{Mongondry1}. Moreover
the understanding of the physical mechanism that can originate an
arrested phase also at clay concentration so low as C$_w$=0.3 $\%$
and with no added salt is particularly intriguing. In our previous
works we have speculated about the origin of the arrested phases
in low and intermediate clay concentration in salt free deionized
water \cite{Ruzicka1,Ruzicka2}. In this paper we want to extend
our measurements performing the same systematic study as a
function of clay concentration but introducing salt in the
solutions to change its ionic strength. In particular we want to
investigate an unexplored region of the phase diagram focusing our
attention at low clay but higher salt concentrations and across
the supposed liquid/gel transition of Mourchid et
al.\cite{Mourchid}. A very recent paper \cite{Li} has investigated
very dilute Laponite solutions at two different salt
concentrations with static light scattering and small-angle x-ray
scattering techniques but at the moment there are not dynamic
light scattering data that are spanning in a complete way this
region of the phase diagram (there are not available data at
$I\geq 10^{-3} M$ \cite{Tanaka2}) and these measurements can be
very useful to understand the intriguing and controversy Laponite
phase diagram.

In this paper we observe that for all the samples studied, for all
clay concentrations (up to very low one, $C_w=0.3 \%$) and all
salt concentrations ($C_s=1 \times 10^{-3} M$, $C_s=2 \times
10^{-3} M$, $C_s=5 \times 10^{-3} M$), the liquid phase is not the
stable one but the samples age toward an arrested phase. These
results confirm the previous ones about the arrested phase found
at very low clay concentrations in pure
water\cite{Ruzicka1,Ruzicka2}. The time necessary for the sample
to evolve in an arrested state depends on clay and salt
concentrations, as we will see later. For none of the samples
studied we have found flocculation, also at the highest studied
salt concentration, in agreement with previous results
\cite{Mourchid, Mongondry1}.
\section{Experimental Section}
\subsection{Materials}
Laponite RD is produced by Laporte Industries with a density of
$2570$ $Kg/m^3$. It is composed of rigid disc-shaped crystals with
a well defined thickness of $1$ nm and an average diameter of
about $30$ nm. Each crystal is composed of roughly $\sim 1500$
unit cells with an empirical chemical formula $Na^+_{0.7}[( Si_8
Mg_{5.5} Li_{0.3})O_{20}(OH)_4]^{-0.7}$. The cell structure is a
layer of six octahedral magnesium ions sandwiched between two
layers of four tetrahedral silicon atoms that are bonded with $Na$
atoms.

When the powder is dispersed in water, the $Na^+$ ions on the
surface are released and a strongly negative charge (roughly 700
elementary charge) appears on the faces. On the other hand,
because of the protonation of the OH groups with the hydrogen
atoms of water, a weakly positive charge appears on the rim of the
discs.
\subsection{Samples Preparation}
Laponite powder was dried at $T=400$ K for 4 hours because up to
20\% of its "as prepared" weight is due to adsorbed water. Then
the powder is dispersed in a solution of NaCl at the desired salt
concentrations $C_s$, prepared by adding salt to MilliQ deionized
water ($18$ $M\Omega$, $pH=7$) in different weight proportions to
obtain $C_s=1 \times 10^{-3} M$, $C_s=2 \times 10^{-3} M$ and
$C_s=5 \times 10^{-3} M$. Laponite dispersion was stirred
vigorously until the sample become transparent and finally
filtered through a 0.45 $\mu m$ pore size Millipore filters. The
starting aging (or waiting) time (t$_w$=0) is defined as the time
when the suspension is filtered.

Chemical dissociation of Laponite platelet is an important
phenomenon, that -if present- could seriously affect the behavior
of the samples. From data available in literature it is in fact
known that Laponite dissociation is actually present in acid
solutions, that could be the result, for example, of leaving
samples in contact with atmospheric CO$_2$. The chemical reaction
of this dissociation is \cite{Thompson, Mourchid2}:
\begin{equation}
Si_8Mg_{5.45}Li_{0.4}H_4O_{24}Na_{0.7}+12 H^++8H_2O \rightarrow 0.7Na^++8Si(OH)_4+5.45Mg^{2+}+0.4Li^+
\end{equation}
It is evident from this reaction that the eventual platelet
dissociation would be signaled by the  presence of $Mg^{++}$ ions
in solution. To avoid this phenomenon, our samples were prepared
with particular care, preventing any contact with air during and
after sample preparation (the samples have been prepared in a
glove box under nitrogen atmosphere, and then sealed in quartz
cells). Another important chemical modification could be due to
the change with aging time of the amount of $Na^+$ ions in
solution. It is in fact well know that at the early stage of the
solution, $Na^+$ ions detaches from the Laponite surfaces and
establish the ionic strength of the solution.

We are confident that no chemical modifications of the solutions
take place after the first 10-20 minutes because after Laponite is
suspended in pure water the pH of the solution, measured by a
Crison Glp 22 pHmeter, reached in few minutes a value in the range
pH=9.8$\div$10.0. The pH was then monitored in part of the sample
left in the glove box under nitrogen atmosphere and we did not
observe any deviance from this value also on a time scale of weeks
or months. Moreover, a set of measurements aiming to determine the
ratio, for both $Na^+$ and $Mg^{++}$, of ions dissolved in the
solution (characterized by a narrow NMR absorption line) to those
stuck in the solid environment of the platelet (broad NMR
absorption band) has been performed on the same samples
\cite{Capuani}. These measurements show that the fraction of
solvated $Na^+$ ions does not change with time, and that the
solvated $Mg^{++}$ ions are absent. Therefore, we are confident
that Laponite dissolution is not present in our samples.

A large number of Laponite dispersions at several salt
concentrations have been prepared: for $C_s=1 \times 10^{-3} M$,
$C_s=2 \times 10^{-3} M$, and $C_s=5\times 10^{-3} M$. The samples
investigated are indicated by dots in the $C_s-C_w$ phase diagram
reported in Figure 1, also indicated are the samples previously
studied \cite{Ruzicka1, Ruzicka2} in pure deionized water ($C_s
\simeq 1 \times 10^{-4} M$). In the following we will name the
samples reported in the figure as open circles as "low"
concentration samples while samples reported as full circles as
"high" concentration samples for the reasons discussed below.
\begin{figure}[h]
\centering
\includegraphics[width=.8\textwidth]{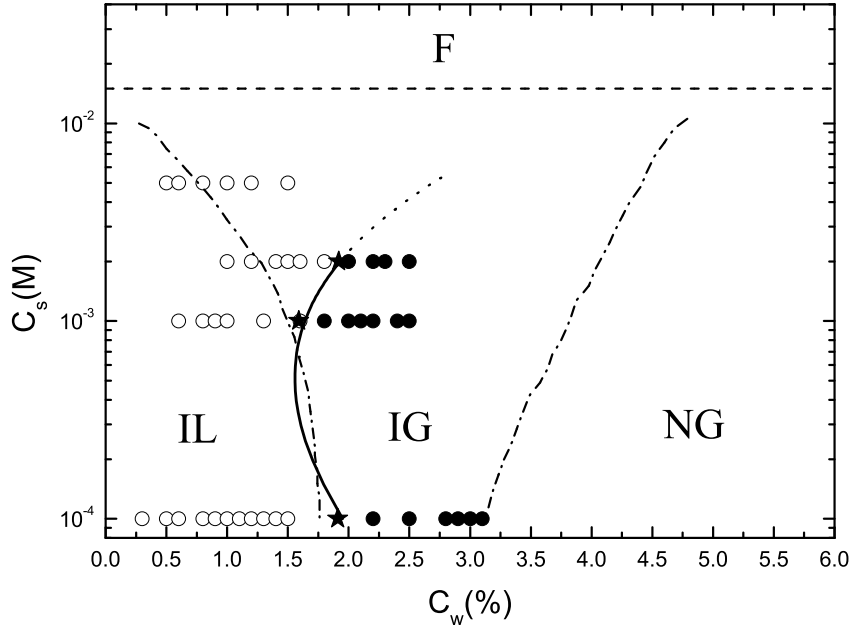}
\caption{$C_s-C_w$ phase diagram proposed in Ref.\cite{Mourchid}.
The isotropic liquid (IL)-isotropic gel (IG) and isotropic gel
(IG)-nematic gel (NG) transition lines from \cite{Mourchid} are
reported as dashed-dotted lines. The dashed line in the upper part
of the diagram separates the flocculation region (F). Open and
full circles indicate respectively "low" and "high" concentration
samples measured in the present work. The stars indicate the
concentration values obtained from the fit analysis of
Fig.~\ref{f4} while the thick full-dotted curve drawn from these
points represents the new line proposed that marks the transition
between the two different aging behaviors. Both at the left and
right of this line the system is arrested and form an isotropic
gel. The last part of the line is dashed because at the highest
studied salt concentration ($C_s=5\times 10^{-3} M$) the aging
dynamics at clay concentration larger than $C_w=1.5$ $\%$ is too
fast to be followed with photoncorrelation measurements. Data at
$C_s=1 \times 10^{-4} M$ are from our previous
works\cite{Ruzicka1, Ruzicka2}.} \label{f1}
\end{figure}

\subsection{Equipment and Measurements}
Dynamic light scattering measurements were performed using an
ALV-5000 logarithmic correlator in combination with a standard
optical setup based on an He-Ne ($\lambda=632,8$ nm) 10 mW laser
and a photomultiplier detector. The intensity correlation function
was directly obtained as:
\begin{equation}
g_2(q,t)=\frac{\langle I(q,t)I(q,0)\rangle}{\langle
I(q,0)\rangle^2},
\end{equation}
where $q$ is the modulus  of the scattering wave vector defined as
$q=(4 \pi n/ \lambda) \sin(\theta /2)$ with $\theta=90^{\circ}$ in
this experiment. The dynamic structure factor $f(q,t)$ can be
directly obtained inverting the Siegert relation :
\begin{equation}
f(q,t)=\sqrt{\frac{g_2(q,t)-1}{b}}.  \label{f(q,t)}
\end{equation}
where $b$ represents the coherence factor.

\section{Results and Discussion}

\subsection{The aging phenomenon in  Laponite suspensions}
According to the phase diagram obtained from Mourchid \textit{et
al.}\cite{Mourchid} up and below a certain concentration $C_w^*$,
that depends on salt concentration $C_s$, Laponite suspensions can
be in two different physical states. Low concentration suspensions
($C_w<C_w^*(C_s)$) should form stable, equilibrium liquid phase
(\textbf{IL} phase of Fig.~\ref{f1}). Higher concentration
suspensions ($C_w>C_w^*(C_s)$) are initially fluids but experience
aging and pass into an arrested phase after a time that depends on
the clay amount and salt concentration (\textbf{IG} phase of
Fig.~\ref{f1}). However, as already discussed, recent
investigations have shown that the phase diagram needs to be
revised. In particular, a surprising arrested phase has been found
also at low clay concentrations (well within the supposed stable
\textbf{IL} phase of Fig.~\ref{f1}) for very long waiting times
\cite{Nicolai1,Nicolai2,Ruzicka1,Ruzicka2,Mongondry1}.

\begin{figure}[h]
\centering
\includegraphics[width=1\textwidth]{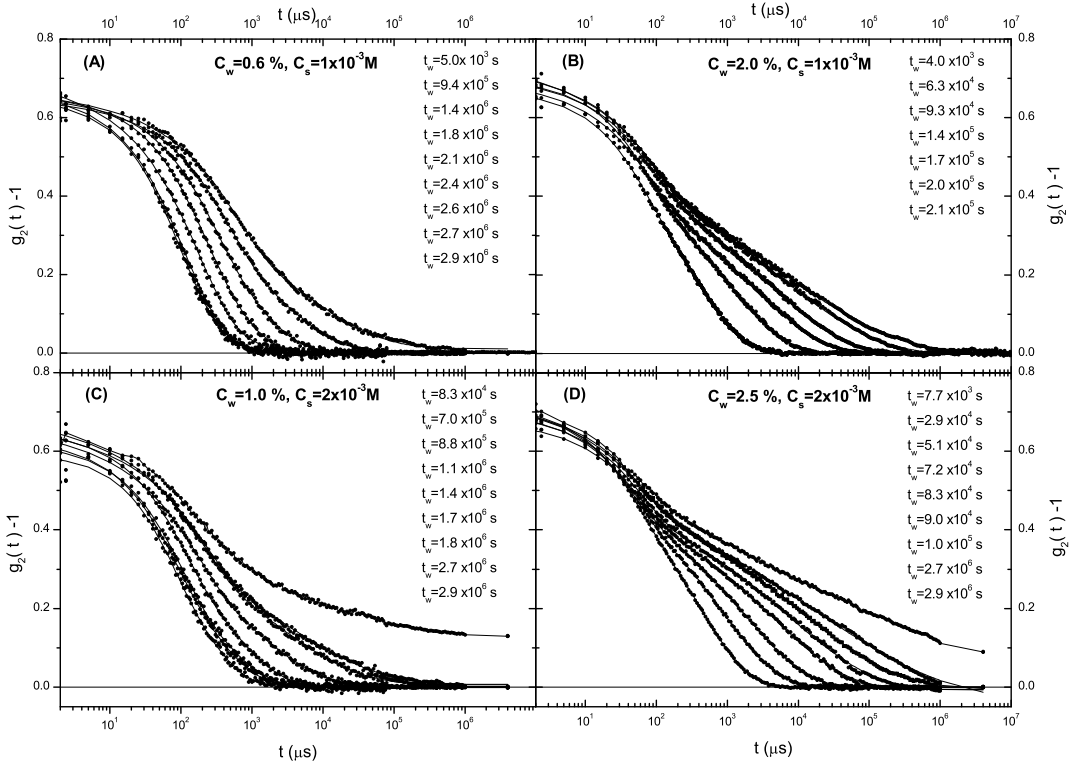}
\caption{Evolution of the measured intensity correlation functions
(symbols) and corresponding fits
  with Eq.~(\ref{fit}) (continuous lines) for four different
  Laponite suspensions $C_w$ at two salt concentrations $C_s$.
  For $C_s=1 \times 10^{-3} M$: $C_w$=0.6 \% (A) and $C_w$=2.0 \%  (B).
  For $C_s=2 \times 10^{-3} M$: $C_w$=1.0 \% (C) and
  $C_w$=2.5  \% (D). The curves are measured at
  increasing waiting times (indicated in the figure) from left to right.
  In panels (C) and (D) also
  the transition in the non ergodic phase are reported.} \label{f2}
\end{figure}

Correlation functions at increasing waiting times t$_w$ for four
samples at two different salt concentrations $C_s$ are reported in
Figure~\ref{f2} as an example. In particular low ($C_w=0.6 \%$ -
panel (A)) and high ($C_w=2.0 \%$ - panel (B)) clay concentration
samples for $C_s=1 \times 10^{-3} M$ and low ($C_w=1.0 \%$ - panel
(C)) and high ($C_w=2.5 \%$ - panel (D)) clay concentration
samples for $C_s=2 \times 10^{-3} M$ are shown. It is evident from
the figure that all the samples are actually performing aging and
that the liquid phase is not the stable one. The evolution of the
spectra with the waiting time (aging) is in fact evident. The
crossover from a complete to incomplete decay of the correlation
function (indication of a transition towards a non ergodic,
arrested, state) is also reported as an example in the lower panel
of Fig.~\ref{f2}. Only the ergodic spectra of each sample have
been studied in the paper. As for the samples prepared in pure
water, the correlation functions (Figure~\ref{f2}) decay following
a two steps behavior (more evident from panels B and D of the
figure) and for this reason we apply to these measurements the
same data analysis used previously \cite{Ruzicka1,Ruzicka2}. The
two different decaying, the fast and the slow ones, can be
described
 quantitatively assuming for $f(q,t)$ a time dependance given by the
 weighted sum of two contributions, a simple
exponential to represent the fast decay (weight $a$) and a
stretched one $[e^{-(t/\tau_2)^{\beta}}]$ for the slow decay
(weight ($1-a$)). The measured quantity, $g_2(q,t)-1$, is thus the
squared sum of the two terms:
\begin{equation}
g_2(q,t)-1=b \left ( a e^{(-t/\tau_1)}+(1-a)
e^{(-(t/\tau_2)^\beta)} \right )^2. \label{fit}
\end{equation}
The fits with Eq.~\ref{fit} well describe the data as can be seen
from the full lines superimposed to the experimental points
reported as an example in Figure~\ref{f2}.

Directly from the raw spectra reported in Fig~\ref{f2} it is
already possible to observe as the evolution with waiting time is
qualitatively different for low and high concentration samples, as
also observed in free salt samples \cite{Ruzicka1, Ruzicka2}.
While in fact for low concentration samples (open circles of
Fig~\ref{f1}) (Fig~\ref{f2}A for $C_s=1 \times 10^{-3} M$ and
Fig~\ref{f2}C for $C_s=2 \times 10^{-3} M$) one observes the
evolution with $t_w$ of the whole $g_2-1$, i.e. both the fast
($\tau_1$) and the slow ($\tau_2$) decays evolve with $t_w$, for
the higher concentration samples (full circles of Fig.~\ref{f1})
(Fig~\ref{f2}B for $C_s=1 \times 10^{-3} M$ and Fig~\ref{f2}D for
$C_s=2 \times 10^{-3} M$) the fast decay remains essentially
constant while the slow one becomes larger and larger when the
waiting time increases. This qualitative observation is fully
confirmed by the data analysis (see below).

\subsection{The slow dynamics: the B parameter and the aggregation rate}

We will focus now our attention on the slow dynamics and its
characteristic parameters.

One can think to the stretched exponential as the results of the
superposition of single exponentials:
\begin{equation}
 e^{-(t/\tau_2)^{\beta}}=\int_0^{\tau} \phi_{\tau_2 , \beta}(\tau ')
 e^{-t/\tau '} d\tau '
\end{equation}
where $\phi_{\tau_2, \beta}(\tau)$ is the appropriate distribution
of relaxation times $\tau$ (this distribution function has not an
expression in terms of elementary functions).

 From the correlation time $\tau_2$ and the stretching
exponent $\beta$ it is then possible to define the mean relaxation
time $\tau_m$ of the distribution $\phi$ that gives rise to the
stretched exponential form:
\begin{equation}
 \tau_m=\int_0^{\tau} \tau ' \phi_{\tau_2 , \beta}(\tau ') d\tau '=\tau_2  \frac{1}{\beta} \Gamma(\frac{1}{\beta}).
 \label{taumeq}
\end{equation}
where $\Gamma$ is the Euler gamma function.

The value of the mean relaxation time $\tau_m$ is larger than
$\tau_2$ and it spans more decades, because of the $t_w$
dependence of $\beta$, but the qualitative behavior of the two
relaxation times is the same. Indeed both $\tau_2$ and $\tau_m$
seem to diverge at a given waiting time $t_{w}^{\infty}$ i.e. when
the system is arrested. This behavior is evident from Fig. 3 of
Ref.\cite{Ruzicka2} where the comparison between $\tau_2$ and
$\tau_m$ for several samples is reported.

\begin{figure}[ht]
\centering
\includegraphics[width=.85\textwidth]{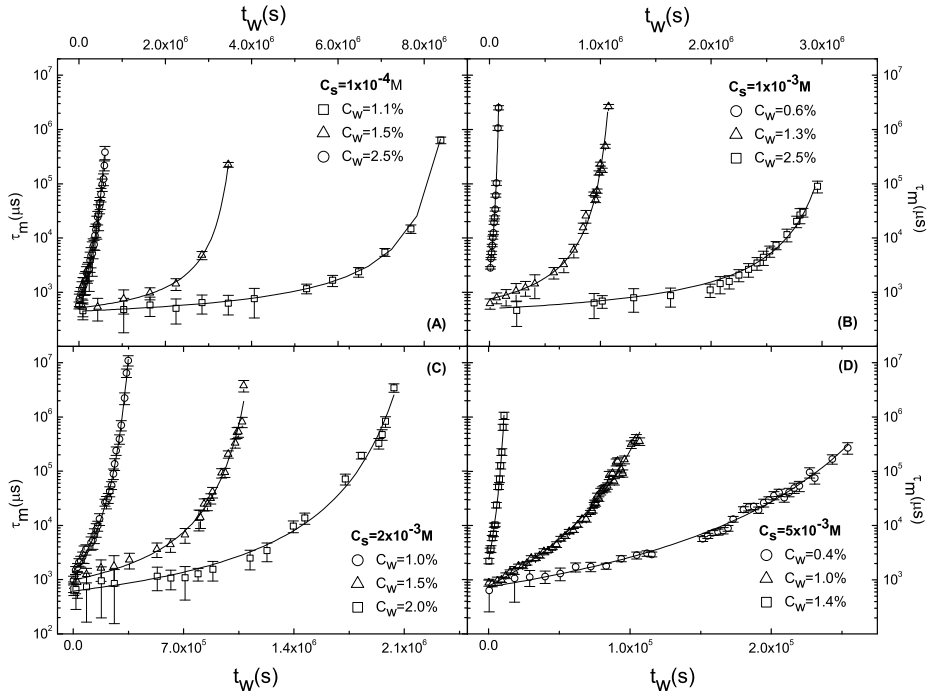}
\caption{Waiting time dependence of $\tau_m$ and relative fits
according to Eq.~\ref{taum} for several sample concentrations
$C_w$ at different salt concentrations $C_s$. (A): $C_w$=1.1; 1.5
and 2.5 \% for $C_s \simeq 1 \times 10^{-4}$ M from
Ref.\cite{Ruzicka1}. (B): $C_w$=0.6; 1.3 and 2.5 \% for $C_s=1
\times 10^{-3}$ M. (C): $C_w$=1.0; 1.5 and 2.0 \% for $C_s=2
\times 10^{-3}$ M. (D):$C_w$= 0.4; 1.0 and 1.4 \% for $C_s=5
\times 10^{-3}$ M.}
 \label{f3}
\end{figure}
In Fig.~\ref{f3} the values of $\tau_m$, calculated according to
Eq.~\ref{taumeq}, for several clay concentrations at the four
studied salt concentrations are reported as examples.

We observe that in the region of times studied in this paper
$\tau_2$ (or equivalently $\tau_m)$ is always much smaller than
$t_w$, this indicating that the measured $g_2$ are well defined
quantities. We clearly observe that the dependence of $\tau_m$ on
the waiting time $t_w$ is faster than exponential. This is not in
contradiction with previous results because the deviation from the
exponential growth is clearly noticeable only at low clay
concentrations, where there are not data available in literature.
At high clay concentration the simple exponential shape can be
confused with our faster than exponential behavior.

 For a quantitative analysis, we represent
the aging time dependence of relaxation times as
\cite{Ruzicka1,Ruzicka2}:
\begin{equation}
 \tau_m = \tau_m^0 \; exp \left (B \frac{t_w}{t_{w}^{\infty}-t_w}
 \right ) \label{taum}
\end{equation}
This equation is not derived from any first principle argument,
rather it embodies the exponential growth of the $\tau_m$ for
short $t_w$ (as previously observed \cite{Bellour,Tanaka2,Kaloun})
and the existence of a divergence of $\tau_m$ at a finite value of
$t_w$, i.e. $t_w^{\infty}$.

Equation~\ref{taum} well describes the measurements as can be seen
from the fit curves reported as continuous lines in Fig~\ref{f3}.

\begin{figure}[h]
\centering
\includegraphics[width=.5\textwidth]{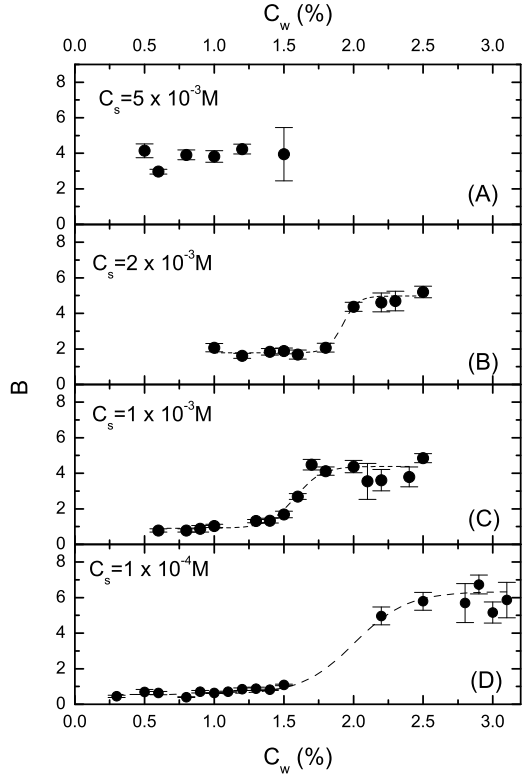}
\caption{Concentration dependence of the $B$ parameter for all the
samples at the indicated salt concentrations decreasing from top
(panel A) to bottom (panel D). Data reported in panel D are from
our previous measurements \cite{Ruzicka1, Ruzicka2}. The
transition from a constant value for the low concentration samples
towards another constant value for the high concentration samples
(reported respectively as open and full circles in Fig~\ref{f1})
is an indication of a transition between two different arrested
states. The line reported in the figure is a sigmoidal fit of the
B parameter.} \label{f4}.
\end{figure}

\begin{figure}[h]
\centering
\includegraphics[width=.5\textwidth]{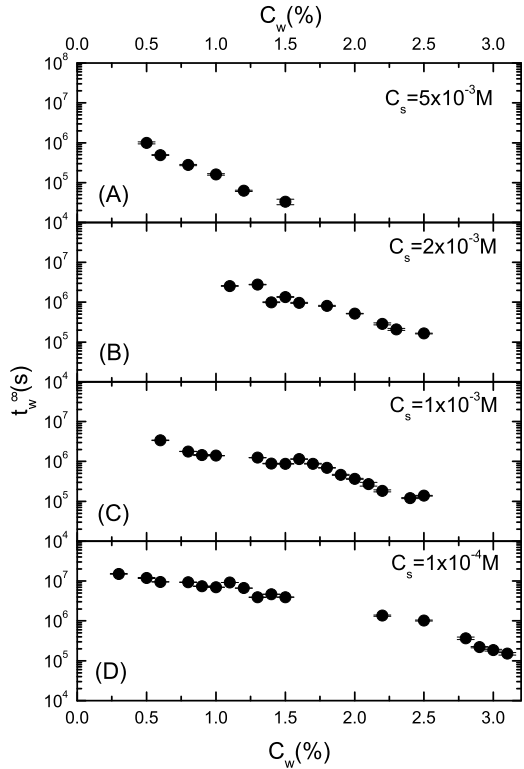}
\caption{Concentration dependence of the divergence time
$t_{w}^{\infty}$ for all the samples at the indicated salt
concentrations decreasing from top (panel A) to bottom (panel D).
Data reported in panel D are from our previous measurements
\cite{Ruzicka1, Ruzicka2}.} \label{f5}
\end{figure}

The parameters of the fits, $B$ and $t_w^{\infty}$, in function of
Laponite concentration $C_w$ at the different salt concentrations
studied are reported in Fig.~\ref{f4} and~\ref{f5}. Salt
concentration is decreasing from the top (A) to the bottom (D) of
the figures where panel D reports our data in pure deionized water
\cite{Ruzicka1,Ruzicka2}.

It is evident from figure~\ref{f4} that the results found
previously for the free salt suspensions are confirmed: the $B$
parameter, that measures how fast $\tau_m$ approaches the
divergence, is almost constant for all the samples at low
concentration and drifts towards a higher, again constant, value
for high concentration samples. This happens for all the series of
studied samples, from the lowest to the highest values of salt
concentrations. In the discussion of Fig.~\ref{f2} we already
observed that the aging process is qualitatively different for the
low and high concentration samples. Figure~\ref{f4} quantify this
difference in the physical properties characterizing the aging
phenomenon in the two different concentration regions. From the
behavior of the B parameter reported in this figure it is evident
that the existence of two different routes toward an arrested
phase found for the samples at $C_s \simeq 1 \times 10^{-4} M$
\cite{Ruzicka1,Ruzicka2} is also present for this new series of
samples with higher ionic strength. The transition between the low
and the high B values should indicate a transition line between
the two different routes to gelation. The position of this line
depends on salt concentration. In Fig.~\ref{f4} a fit of the B
parameter for each salt concentration with a sigmoid curve is also
reported. From this fit analysis is possible to extract the mean
concentration value of the transition, and this gives an
indication of the transition line. These values are reported as
 stars in Fig.~\ref{f1}

In Fig.~\ref{f5} the behavior of the $t_{w}^{\infty}$ parameter in
function of the clay concentration $C_w$ for different salt values
is reported. We can observe that for each sample at each salt
concentration the $t_{w}^{\infty}$ parameter actually exists and
is finite; this is the indication that all the studied samples are
not in a stable phase but evolve with waiting time toward an
arrested state. The value of $t_{w}^{\infty}$ changes with clay
and salt concentration. At fixed salt concentration $C_s$ it
decreases at increasing Laponite concentration $C_w$, in agreement
with the fact that more the sample is concentrated faster is the
aggregation process. $t_{w}^{\infty}$ generally decreases also
with increasing salt concentration $C_s$ (passing from panel D to
panel A), in particular it is strongly decreased going from $C_s
\simeq 1 \times 10^{-4} M$ to $C_s=1 \times 10^{-3} M$, it is
essentially constant going to $C_s=2 \times 10^{-3} M$, and it is
again strongly reduced for $C_s=5 \times 10^{-3} M$. Also this
behavior is in agreement with the fact that the aggregation is
faster for the samples with higher salt concentration. It must be
emphasized that $t_w^{\infty}$ is smooth and continuous with
$C_w$, and it does not show any signature of the transition
concentration, where $B$ is discontinuous.

\subsection{The hypothesized phase diagram for low concentrations
sample} From the measurements reported in this paper and from that
reported in free salt water \cite{Ruzicka1,Ruzicka2} we can now
draw an alternative phase diagram for Laponite solutions. At
variance with the results of \cite{Mourchid, Tanaka1} it is
definitely clear that the low clay concentration region of the
phase diagram is neither a liquid neither a sol stable phase, both
at low salt concentration ($C_s \simeq 10^{-4} M$) and at higher
salt concentration. The state of the system depends on time and
 the very long waiting time necessary to
obtain the arrested phase (especially at low salt low clay
concentrations) can be the reason because previous studies have
believed the initial liquid phase to be the stable one. In regard
of this point we can observe  from Fig.~\ref{f5} that for all the
different salt concentrations the parameter $t_{w}^{\infty}$
increases in a continuous way as clay concentration is decreased
(as already discussed). Therefore we believe that also the samples
found in a liquid phase from Ref.~\cite{Mongondry1}, will transit
in an arrested phase for a long enough waiting time. This time can
be obtained by an extrapolation at the interested clay
concentration of the $t_{w}^{\infty}$ reported in Fig.~\ref{f5}
and, for example, for $C_s=1 \times 10^{-3} M$ it results to be
around 3 months or more for a clay concentration of $C_w=0.3$ \%.
However, due to the lack of information on this low $C_w$ region,
we cannot exclude the possibility that at low enough concentration
a really stable liquid phase actually exists.

From the behavior of the B parameter reported in Fig.~\ref{f4} we
can also state that in the high clay concentration region no
discontinuity is found increasing salt concentration from $C_s=1
\times 10^{-4} M$ up to $C_s=5 \times 10^{-3} M$ while a general
slow decrease of the "high" value of the B parameter is found.
This is an important issue that enters in the discussion about the
origin of the arrested phase at $C_s=1 \times 10^{-4} M$ as glass
or gel phase \cite{Tanaka1, Mongondry1}. Our measurements are of
dynamic type so we cannot access the structure factor of the
samples and we cannot distinguish in this sense between the gel
and the glass phase but we can say that there is not any
discontinuity between  $C_s=1 \times 10^{-4} M$ and higher salt
concentrations. For this reason the arrested phase existing in
free salt water should persist also adding salt to the solutions.
Relying on the fact that for $C_s
>1 \times 10^{-3} M$ the origin of the arrested phase is universally
assigned to gelation \cite{Tanaka1, Mongondry1}, we conclude that
the same gel phase should be present at $C_s=1 \times 10^{-4} M$.

Also the low clay concentration phase (low values of the B
parameters) has no signs of discontinuity, nevertheless the "low"
value of the B parameter increases at increasing salt
concentration.

For these reasons moving vertically in the phase diagram, i.e.
fixing a clay concentration value and increasing salt
concentration, one finds a continuous evolution of the B parameter
both at low and at high clay concentrations.

The behavior is completely different if we span horizontally the
phase diagram, i.e. if we fix salt concentration and increase clay
concentration. For all the different salt concentration values we
find a discontinuity in the B value and a drift from a "low" to a
"high" value of this parameter. Therefore we have found again, as
already seen in free salt water, the existence of two different
routes to reach the arrested phase, for low and high clay
concentration. The fit analysis of Fig~\ref{f4} permits also to
obtain an estimation of the concentration value that represents
the transition point between the two different arrested states. In
this way we can draw a new transition line in the phase diagram
between two different arrested phases. This is reported as
full-dotted line in Fig.~\ref{f1}.

As already said before we cannot have a direct evidence of the
gel/glass origin of the non ergodic phases but we can speculate
that the arrested phase at low clay concentration is originated
from clusters of Laponite while single Laponite disks directly
participate to the formation of the arrested phase at higher clay
concentration.
\section{Conclusions}
From our previous work about Laponite dispersions in salt free
deionized water we have shown the presence of an arrested phase at
very low clay concentration and the existence of two different
routes to reach the aggregated phase for "low" and "high"  clay
concentration \cite{Ruzicka1,Ruzicka2}. In this paper we have
investigated the effect of increasing the ionic strength of a
water dispersion of Laponite. Systematic measurements at three
different values of salt concentrations have been in fact
performed and these data give now information on a wide part of
Laponite solutions phase diagram. The results obtained have
undoubtedly confirmed the existence of an arrested phase at low
clay concentration for each salt concentration. Moreover the
measurements have also validated the goodness of the proposed
analysis for samples in pure deionized water and the existence of
two different aggregation mechanisms. In particular at the fixed
clay concentration $C_w=3$ $\%$ no discontinuity is found
increasing salt concentration, thus confirming the hypothesis of a
gel phase in the whole ionic strength range proposed in
Ref.\cite{Mongondry1}. On the contrary at each salt concentration
investigated here, increasing clay concentration a transition line
between two different aggregated phases is found.

\end{document}